\documentclass[twocolumn,superscriptaddress,nofootinbib,pre]{revtex4-1}
\usepackage[dvipdfmx]{graphicx}
\usepackage{epsfig}
\usepackage{amssymb}
\usepackage{amsmath}
\usepackage{amsbsy}
\usepackage{color}
\usepackage{ulem}
\usepackage{epstopdf}

\begin{document}
\title{Note on two-point mean square displacement}
\author{Naoya Katayama}
\author{Takahiro Sakaue}
\affiliation{Department of Physical Sciences, Aoyama Gakuin University, 5-10-1 Fuchinobe, Chuo-ku, Sagamihara, Japan}
\thanks{corresponding author, sakaue@phys.aoyama.ac.jp}

\begin{abstract}
When probe molecules of interest are embedded in a substrate or aggregate under stochastic motion, one needs to rely on the so-called two-point mean square displacement (MSD) measurement to extract the intrinsic mobility of the probes. We discuss two versions, based on the time series of {\it relative vector} or  {\it distance} between two probes, and summarize their basic properties compared to the standard MSD. We also propose a way to extract (i) the non-Gaussianity in the displacement statistics and (ii) the motional correlation between probes from the two-point MSD. The results are presented not only for independent probes, but also for intramolecular probes within a long polymer, which could be useful in quantifying the dynamics of chromatin loci in living cell nucleus.
\end{abstract}

\maketitle
\section{Introduction}
Single particle tracking is a method of analysis widely used in material and biological sciences~\cite{Qian1991, Metzler2014}. By attaching a fluorescent tag to the molecule of interest, it enables us to observe and record its real time motion under an optical microscope ~\cite{Shen2017, Clarke2019}. The motional characteristics thus obtained are expected to carry a rich information on the system. 
In typical applications,  one calculates from the stochastic trajectory ${\vec r}(t)$ the mean-square displacement (MSD);
\begin{eqnarray}
{\rm MSD}(\tau) \equiv \langle | \Delta {\vec r}( \tau) |^2\rangle 
\label{MSD}
\end{eqnarray}
where $ \Delta {\vec r}( \tau) = {\vec r}(t_0 + \tau) - {\vec r}(t_0)$. MSD thus measures the typical displacement squared during the time interval $\tau$. Note that MSD generally depends on the time origin $t_0$. In the present note, we assume stationary state, where time translational invariance implies no dependence of statistical quantities on $t_0$, and $\langle \cdots \rangle$ represents the average over trajectories and over the time origin $t_0$.

\begin{figure}[h]
	\centering
	\includegraphics[width=0.3\textwidth]{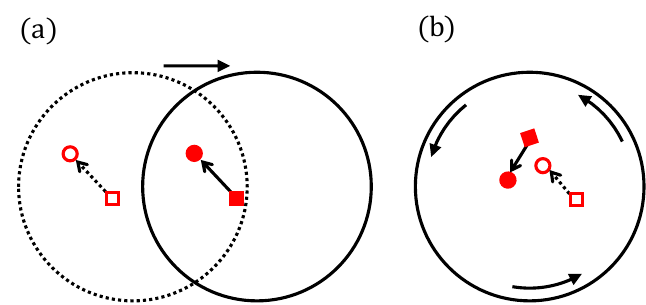}
	\caption{Schematics illustrating the effect of (a) translational or (b) rotational motion of a substrate on the relative vector ${\vec d(t)}$ and the distance $d(t)$. Intrinsic mobility of probes are hypothetically set to zero to emphasize the extrinsic effect. }
	\label{fig1}
			\vspace{0.2 cm}
\end{figure}

\begin{figure}[h]
	\centering
	\includegraphics[width=0.4\textwidth]{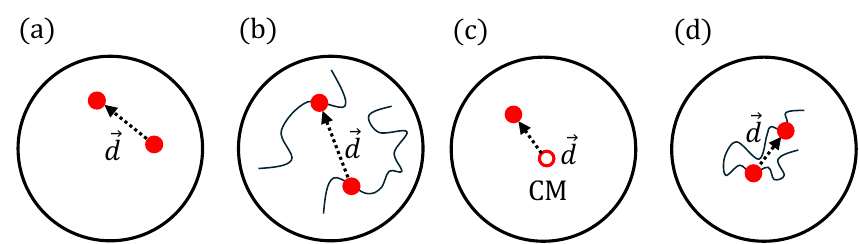}
	\caption{Possible variations for a pair of probes in two-point MSD measurement.  (a) and (b): Pair of independent probes; while probes are simple particles in (a), tagged monomers play a role of probe in (b), each of which belongs to different polymers. (c) Instead of the second probe, one can make use of the center of mass position ${\vec r_c}(t)$ (or other fixed point) in the moving substrate to define a vector ${\vec d}(t) = {\vec r}_1(t) - {\vec r}_c(t)$. (d) Intramolecular probe pair as an example of mutually dependent probes.   }
	\label{fig2}
			\vspace{0.2 cm}
\end{figure}

Although calculation of MSD using Eq.~(\ref{MSD}) looks quite straightforward, one may encounter situations, often in the measurement in biological samples, where it is not the case. Indeed, a complication arises when the molecule of interest is embedded in a medium or an aggregate, hereafter called substrate, which itself moves. In such a case, the observed displacement $\Delta {\vec r}^{(o)}(\tau)$ is a superposition of two contributions;
\begin{eqnarray}
    \Delta {\vec r}^{(o)}(\tau) = \Delta {\vec r}(\tau) + \Delta {\vec r}^{(ex)}(\tau)
\label{delta_r_o}
\end{eqnarray}
where $\Delta {\vec r}(\tau)$ represents the displacement due to the intrinsic dynamics of probe, while $\Delta {\vec r}^{(ex)}(\tau)$ stands for the extrinsic contribution arising from the substrate motion. If one uses $\Delta {\vec r}^{(o)}(\tau)$ in Eq.~(\ref{MSD}), the resultant {\it observed} MSD, which we denote as ${\rm MSD}^{(o)}(\tau)$, would be different from ${\rm MSD}(\tau)$, thus fails to capture the intrinsic mobility of the probe.
The best example is found in the analysis of chromatin dynamics inside a cell nucleus~\cite{Marshall1997, Mine-Hattab2012, Ochiai2015, Arai2017, Khanna2019, Yesbolatova2022, Gabriele2022, Bruckner2023, Nozaki2023, Minami2024}; we are interested in the motion of chromatin locus, but the time series of its position reflects not only the intrinsic chromatin dynamics, but also the translational and rotational motion of the nucleus, which is subjected to the random or systematic forces from cytoplasm, see Fig.~\ref{fig1}. While such an effect of the nuclear motion might be negligible in short time scale, it becomes increasingly visible in longer time scale. For instance, in early embryos of {\it C. elegans}, the nuclear motion dominates the apparent motion of chromatin locus already at $ \tau \sim 10 $ seconds~\cite{Yesbolatova2022}. The relevance of nuclear motion has also been reported in chromatin dynamics in mouse embryonic stem cells~\cite{Oliveira2021}.  A natural strategy here to extract the intrinsic chromatin mobility is to make use of the trajectories of two probes. 
The same problem is encountered in many other cases, for instance, when analyzing the motion of individual cells within multicellular aggregates. With such a situation in mind, P\"onisch and Zaburdaev have recently studied the statistics of the relative distance between independently moving Brownian particles~\cite{Ponisch2018}. In the chromatin problem, the polymeric effect should be taken into account, in which one may look at not only independently moving intermolecular pair but also intramolecular pair, see Fig.~\ref{fig2}~\cite{Gabriele2022, Bruckner2023, Nozaki2023, Minami2024}. 

Let ${\vec r}_1(t)$ and  ${\vec r}_2(t)$ to represent the positions of the two probes at time $t$. The method of analysis using ${\vec r}_1(t)$ and  ${\vec r}_2(t)$ to quantify the intrinsic mobility of probe may be called {\it two-point MSD} analysis. We note that the term two-point MSD has so far been used in a loose sense without a proper definition.  In literature, there are two variants of the two-point MSD analysis, which differ in the definition of the analyzed quantities, i.e., one may look at the time series of either a relative vector ${\vec d}(t) = {\vec r}_1(t) - {\vec r}_2(t)$ connecting  two probes or its absolute value $d(t) = |{\vec d}(t)|$, the latter being a distance between the probes.
In this note, we aim to clarify the similarity and difference between these two analyses, and their quantitative connection to the standard MSD. 
In Sec.~\ref{sec:short_summary}, we first define the quantities considered in two-point MSD analysis, and present a short summary of the paper. We then proceed to the analysis of simple solvable models, first the case with two independently moving probes in Sec.~\ref{sec:Independent-probes} and then two probes belonging to the common polymer, i.e., two loci in the same chromatin in Sec.~\ref{sec:Intramolecular-probes}. Sec.~\ref{sec:DIscussions} is devoted to discussion on the general aspect of two-point MSD. We also demonstrate that the method can be applied to quantify the non-Gaussian parameter in the displacement statistics and the motional correlation of probe pair.
\section{Short summary}
\label{sec:short_summary}
In this section, we first define the central quantities analyzed in two-point MSD analysis, and present a short summary of the present paper. More detailed analyses and discussions can be found in subsequent sections. Readers who are interested only in the general aspects of two-point MSD analysis may skip Secs.~\ref{sec:Independent-probes} and~\ref{sec:Intramolecular-probes}, where detailed analyses on simple solvable models are presented.

\subsection{Quantities analyzed in two-point MSD}
\label{sec:Quantities}
We define the mean-square change in the relative vector ${\vec d}$ (MSCV) as
\begin{eqnarray}
{\rm MSCV}(\tau) \equiv \langle | \Delta {\vec d}( \tau)|^2\rangle
\label{MSCR1}
\end{eqnarray}
with $\Delta {\vec d}(\tau) =  {\vec d}(t_0 + \tau) - {\vec d}(t_0) $.
Similarly, we define the mean-square change in the distance $d = |{\vec d}|$ (MSCD) as
\begin{eqnarray}
{\rm MSCD}(\tau) \equiv \langle [\Delta d( \tau)]^2\rangle
\label{MSCD1}
\end{eqnarray}
with $\Delta  d(\tau) =   d(t_0 + \tau) - d(t_0)$.
One can rewrite the change in the vector as $\Delta {\vec d}(\tau)= \Delta {\vec r}_1(\tau) - \Delta {\vec r}_2(\tau)$, which indicates that $\Delta {\vec d}(\tau)$ may be also influenced by the substrate motion. Therefore, similarly to ${\rm MSD}^{(o)}(\tau)$, we also define ${\rm MSCV}^{(o)}(\tau)$ and ${\rm MSCD}^{(o)}(\tau)$, which are obtained from the observed displacement $\Delta {\vec r}_i^{(o)}(\tau)$ for $i=1, 2$.

An important difference between these quantities is found by observing that MSCV excludes only the translational, but not the rotational motion of the substrate, i.e., ${\rm MSCV}^{(o)}(\tau) = {\rm MSCV}(\tau)$ only if  the rotational motion of the substrate is negligible see Fig.~\ref{fig1}. On the other hand, the equality ${\rm MSCD}^{(o)}(\tau) = {\rm MSCD}(\tau)$ always holds irrespective of the type of substrate motion.

\begin{figure}[h]
	\centering
	\includegraphics[width=0.4\textwidth]{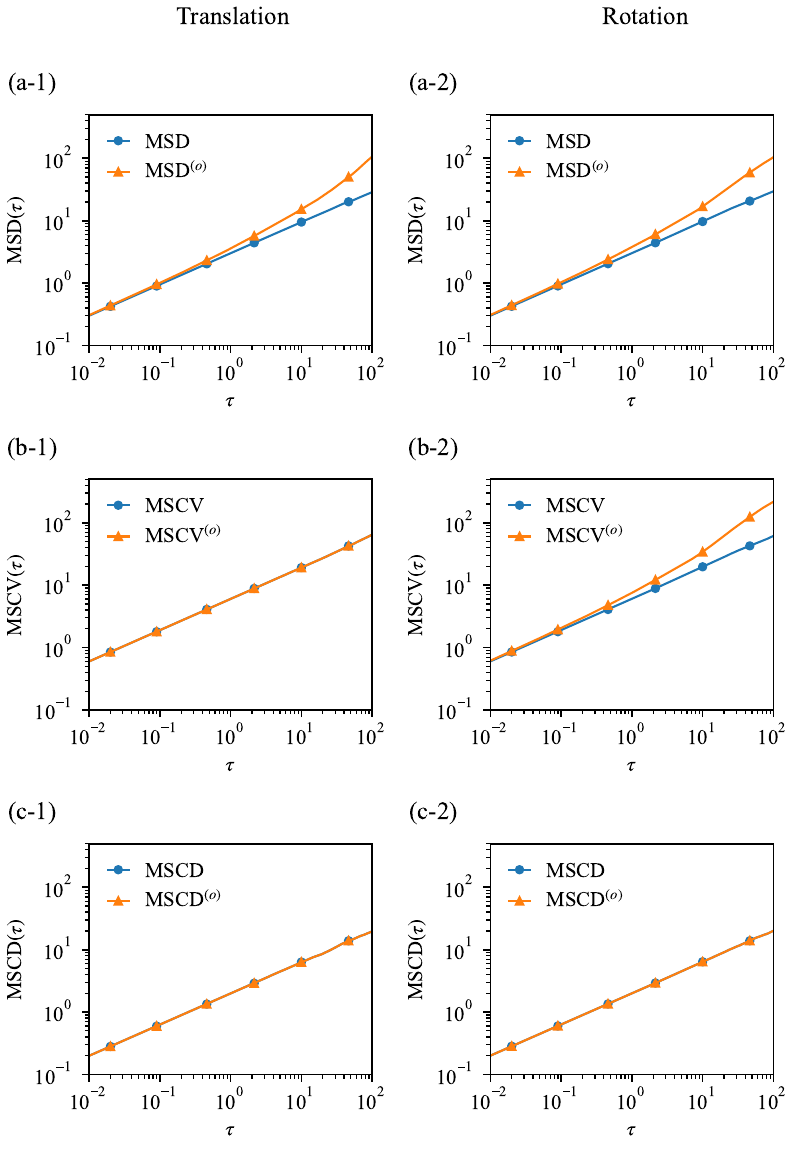}
	\caption{(a)MSD, (b)MSCV and (c)MSCD of probe (or probe pair) performing fractional Brownian motion with MSD exponent $\alpha=0.5$ under the influence of stochastic substrate motion; (left) translational diffusion and (right) rotational diffusion. In each case, both the apparent and the intrinsic quantities are plotted. The former is calculated from the observed displacement $\Delta {\vec r}^{(o)}(\tau)$ and marked by the superscript $(o)$, see the main text for details. Length is measured in unit of probe size, and the unit of time is set by the corresponding time-scale, i.e., the time interval for the probe to make displacement over its own size, see Appendix~\ref{App2-1} for details of the model.    }
	\label{fig3}
			\vspace{0.2 cm}
\end{figure}

As a demonstration, we have carried out a simple simulation for probe particles in a moving substrate (see Appendix~\ref{App2-1} for simulation details). 
As a model of the intrinsic probe motion, we have adopted the sub-diffusional fractional Brownian motion, which mimics the motion of a tagged monomer in polymer.  In Fig.~\ref{fig3} (a), we show results of MSD calculated from $\Delta {\vec r}(\tau)$ or $\Delta{\vec r}^{(o)}(\tau)$, where $\Delta {\vec r}^{(o)}(\tau)$ includes the contribution from the substrate translational (left) or rotational (right) diffusive motion. In both cases, while the substrate motion is negligible in a short time scale, it eventually dominates over the intrinsic mobility in a longer time scale, leading to a spurious higher MSD exponent. It is expected that the more vigorous the substrate motion is, the shorter this crossover time scale (see Appendix~\ref{App2-1} for the estimation of the time scale).   Figs.~\ref{fig3} (b) and (c) show the same analysis for MSCV and MSCD, respectively. Clearly, while MSCV is capable of eliminating the substrate translational motion only, MSCD succeeds in quantifying the intrinsic probe mobility embedded in a substrate undergoing translational and rotational motions. 

\subsection{Relation between MSD, MSCV and MSCD}
\label{sec:relation_MSD}

Given the uncontrollable dependence of ${\rm MSD}^{(o)}$ on the substrate motion, and the ability of its partial or complete removal in two-point MSD analysis, it is useful to find a formula to obtain MSD from MSCD or MSCV. If probes are uncorrelated in their motion, such a formula is given by
\begin{eqnarray} 
 {\rm MSD}(\tau)= \frac{1}{2} {\rm MSCV}(\tau) = \frac{3}{2}  {\rm MSCD}(\tau)  \ 
\label{MSCD_MSD}
\end{eqnarray}
where the mobilities of a pair of two probes are assumed to be equal.
One can provide simple and intuitive explanations for the relation. First, ${\rm MSCV} = {\rm MSD}_1 + {\rm MSD}_2$, where ${\rm MSD}_i$ is the MSD of $i$-th probe, as long as two probes are uncorrelated. It follows directly from the fact that variances of uncorrelated random variables add.
Second,  ${\rm MSCD} = {\rm MSCV}/3 = ({\rm MSD}_1 + {\rm MSD}_2)/3$ follows from the fact that changes in the vector ${\vec d} = {\vec r}_1 - {\vec r}_2$ can be decomposed into radial (1 dimension) and tangential (3-1=2 dimensions), where MSCD is sensitive only to the radial component in {\it short time scale}. The precise meaning of the short time scale, hence the applicability of Eq.~(\ref{MSCD_MSD}) for MSCD will be clarified in the following sections using simple concrete models in Secs.~\ref{sec:Independent-probes} and~\ref{sec:Intramolecular-probes}. A more general discussion is given in Sec.~\ref{sec:DIscussions}.

\section{Independently moving probes}
\label{sec:Independent-probes}
In this section, we present a detailed two-point MSD analysis for the simplest model situation, where probe particles perform independent Brownian motion. The case of intramolecular probe pair within a polymer will be treated in Sec.~\ref{sec:Intramolecular-probes}. 
\subsection{Relative vector analysis}
\label{sec:Independent-probes_MSCV}
\begin{figure}[h]
	\centering
	\includegraphics[width=0.25\textwidth]{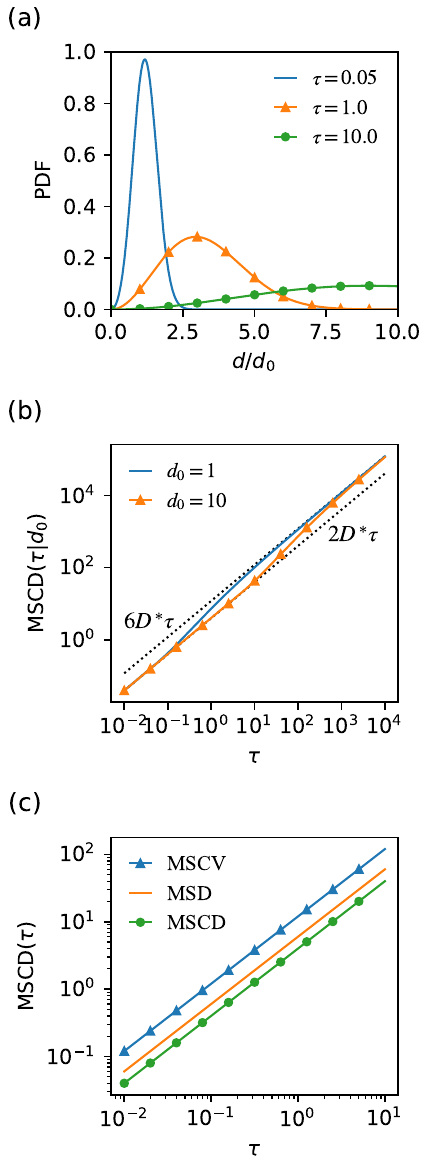}
	\caption{Two-point MSD analysis for a pair of independent probes.  (a) Time evolution of the probability density function $P_d(d,t | d_{0})$ of the inter-probe distance $d$ starting from $d(0)=d_0$. (b) Conditional $ {\rm MSCD}(\tau|d_0)$ as a function of lag time $\tau$ for two different initial separation $d_0$. (c)$  {\rm MSCV}(\tau)$ and ${\rm MSCD}(\tau)$ as a function of lag time $\tau$ compared to ${\rm MSD}(\tau)$. In (b) and (c), the length is measured in unit of $a$ (probe diameter) and the time is measured in unit of $a^2/2D^*$. }
	\label{fig4}
\end{figure}

Consider probes performing Brownian motion in three dimensional space. The probability density function of position ${\vec r}_i$ of each probe ($i=1, 2$) with diffusion coefficient $D_i$ and initial position ${\vec r}_{i0}$ is
\begin{eqnarray}
P_i({\vec r}_i, \tau| {\vec r}_{i0}) = \frac{1}{(4 \pi D_i \tau)^{3/2}} \exp{\left(- \frac{|{\vec r}_i - {\vec r}_{i0}|^2}{4 D_i \tau} \right)}
\label{eq:P_i}
\end{eqnarray}
Then,  MSD of each probe is
\begin{eqnarray}
{\rm MSD}_i(\tau) = \int \mathrm{d} {\vec r}_i \   |\Delta {\vec r}_i(\tau)|^2 P_i({\vec r}_i, \tau | {\vec r}_{i0}) =  6 D_i \tau
\end{eqnarray}
Without loss of generality, we set ${\vec r}_{10}=(0,0,d_0), {\vec r}_{20}=(0,0,0)$.
From Eq.~(\ref{eq:P_i}), the probability density function $P_{{\vec d}}({\vec d}, \tau)$ of relative vector ${\vec d} = {\vec r}_1 - {\vec r}_2 = (d_x, d_y, d_z)$ is obtained as
\begin{eqnarray}
&&P_{{\vec d}}({\vec d}, \tau | {\vec d}_{0}) = \int \mathrm{d} {\vec r}_2 P_1({\vec r}_2+ {\vec d}, \tau | {\vec r}_{10}) P_2({\vec r}_2 , \tau | {\vec r}_{20}) \nonumber \\
&=& \frac{1}{(4 \pi D^* \tau)^{3/2}}\exp{\left( - \frac{d_x^2 + d_y^2 + (d_z-d_0)^2}{4 D^* \tau}\right)}
\label{eq:P_r}
\end{eqnarray}
This is a well-known result, i.e., in the rest frame of probe 2, the motion of probe 1 looks as Brownian motion with diffusion coefficient $D^*= D_1 + D_2$~\cite{Smoluchowski1917, Gillespie}, where, for generality of argument, we introduce the relative diffusibity $D^* $. It then follows 
\begin{eqnarray}
{\rm MSCV}(\tau) = \int \mathrm{d}{\vec d} \   |\Delta {\vec d}|^2 P_{{\vec d}}({\vec d}, \tau | {\vec d}_{0}) =  6 D^* \tau.
\label{MSCV_1}
\end{eqnarray}
which results in twice the MSD if $D_1 = D_2$, i.e., ${\rm MSCV}(\tau) = 2 {\rm MSD} (\tau)$ for two independent identical probes, see Fig.~\ref{fig4} (c).

\subsection{Distance analysis}
\label{sec:Independent-probes_MSCD}
P\"onisch and Zaburdaev have presented a detailed analysis on the statistics of distance $d$ between two independent Brownian particles in two dimension~\cite{Ponisch2018}. Following their analysis, here we present the result in three dimension. First, we transform the probability density function~(\ref{eq:P_r}) of relative vector ${\vec d}$ to that of distance $d$;
\begin{eqnarray}
&&P_d(d,\tau | d_{0}) = \int \mathrm{d} {\vec d} \ P_{{\vec d}}({\vec d},\tau | {\vec d}_{0}) \delta \left(d - \sqrt{d_x^2 + d_y^2 + d_z^2} \right) \nonumber \\
&=& \frac{d}{d_0\sqrt{4 \pi D^* \tau}}  \left[ \exp{\left(-\frac{(d-d_0)^2}{4 D^* \tau} \right)} -  \exp{\left( -\frac{(d+d_0)^2}{4 D^* \tau} \right)}\right] \nonumber\\
\label{P_d_indep}
\end{eqnarray}

The time evolution of $P_d(d,\tau | d_{0})$ is shown in Fig.~\ref{fig4} (a), where the width of the distribution grows as $\sim \tau^{1/2}$.
From Eq.~(\ref{P_d_indep}), we calculate the MSCD under the condition $d(0)=d_0$;
\begin{eqnarray}
 {\rm MSCD} (\tau| d_0)  &\equiv& \int_0^{\infty} \mathrm{d}d \ (d - d_0)^2 P_d(d,\tau | d_{0})  \nonumber \\
&=& 6D^* \tau + 2d_0^2 - 4 d_0 \sqrt{\frac{D^* \tau}{\pi}} \exp{\left(-\frac{d_0^2}{4D^* \tau}\right) }  \nonumber \\
&&- 2(2 D^* \tau + d_0^2) \mathrm{erf}\left( \frac{d_0}{\sqrt{4 D^* \tau}}\right)
\label{MSCD_d0_indep}
\end{eqnarray}
Unlike ${\rm MSD}$ and ${\rm MSCV}$, this quantity depends explicitly on the initial condition,
and this introduces the $d_0$ dependent time scale $\tau_d = d_0^2/4D^*$. With the asymptotic behavior of the error function, we find
\begin{eqnarray}
 {\rm MSCD} (\tau| d_0) \simeq \left\{
 \begin{array}{ll}
 2 D^* \tau & ( \tau \ll \tau_d) \\
 6 D^* \tau & (\tau \gg \tau_d)
 \end{array}
\label{MSCD_d}
 \right.
 \end{eqnarray}
Figure.~\ref{fig4} (b) summarizes the above result, where we plot how $ {\rm MSCD} (\tau; d_0) = \langle (d(\tau)-d_0)^2 \rangle_{d_0}$ evolves with time scale. We note that Eq.~(\ref{MSCD_d}) is consistent with the result of Yesbolatova et al.~\cite{Yesbolatova2022} and that of P\"onisch and Zaburdaev~\cite{Ponisch2018}. In the latter work, $ 6 D^* \tau$  in long time-scale is replaced with $ 4 D^* \tau$ in two dimensional systems.

From experimental standpoint, it would not be easy to take statistics with fixed $d_0$. In typical experiments, one usually takes time averaging, where the time origin, and corresponding $d_0$ continuously change. With the distribution $p(d_0)$ of the initial separation $d_0$ during measurement, one thus defines
\begin{eqnarray}
{\rm MSCD}(\tau) \equiv \int_0^{\infty} \ {\mathrm d} d_0 \ {\rm MSCD} (\tau| d_0)  p(d_0)
\label{MSCD_ta}
\end{eqnarray}
Using Eq.~(\ref{MSCD_d}), we can evaluate ${\rm MSCD}(\tau)$ as
\begin{eqnarray}
{\rm MSCD}(\tau) \simeq q_{>} \times 2 D^* \tau + q_{<} \times 6 D^* \tau 
\end{eqnarray}
with
\begin{eqnarray}
 q_{>} = \int_{\sqrt{4D^*\tau}}^{d_m}\ p(d_0) \  {\mathrm d} d_0 
\end{eqnarray} 
and $ q_{<} =1 - q_{>}$, where we introduce the maximum $d_{m}$ for the allowed $d_0$.
For a pair of free probes, the relative vector ${\vec d}$ is uniformly sampled, thus
\begin{eqnarray}
p(d_0) = \frac{4 \pi d_0^2}{4 \pi d_m^3/3}=\frac{3 d_0^2}{d_{m}^3}
\end{eqnarray}
At short time scale $\tau \ll d_m^2/4D^*$, one then finds $q_{>} \rightarrow 1$, hence, 
\begin{eqnarray}
{\rm MSCD}(\tau)=  2 D^* \tau 
\label{MSCD_1}
\end{eqnarray}
for two independently moving probes, see Fig.~\ref{fig2} (a). In this short time scale regime, the particles move much less than their average initial separation. This regime grows with $d_m$ and extends indefinitely if the particles are on average infinitely far apart. If two probes are identical, we find ${\rm MSCD}(\tau)=  4 D \tau = 2/3 {\rm MSD}(\tau)$.

 \subsection{Motion analysis in center-of-mass frame}
 \label{sec:CM_frame}
 Instead of tracking two probes, one can track the position of a single probe ${\vec r}_1(t)$ with respect to a particular fixed point in the moving substrate. A convenient choice for the fixed point would be the center of mass, whose position ${\vec r}_c(t)$ dynamically changes with the substrate translation~\cite{Ochiai2015}. Similar to the two-probes tracking case, one can define the vector ${\vec d}(t) = {\vec r}_1(t) - {\vec r}_c(t)$, and its magnitude $d(t) = |{\vec d}(t)|$, see Fig.~\ref{fig2} (c).
 In order to avoid a possible confusion, let us attach a subscript ``c"  as ${\rm MSCV_c}$ and ${\rm MSCD_c}$ to indicate the quantities of interest in the center-of-mass frame analysis. 
 Since the diffusivity of the fixed point (second probe) is obviously zero, the relative diffusion coefficient in this case is simply $D^* = D_1$. Thus, we find from Eq.~(\ref{MSCV_1})
 \begin{eqnarray}
 {\rm MSCV_c}(\tau) = 6 D_1 \tau = {\rm MSD}(\tau) = \frac{1}{2} {\rm MSCV}(\tau) , 
  \label{MSCV_c}
 \end{eqnarray}
 and from  Eq.~(\ref{MSCD_1})
  \begin{eqnarray}
 {\rm MSCD_c}(\tau) = 2 D_1 \tau = \frac{1}{3}{\rm MSD}(\tau) = \frac{1}{2} {\rm MSCD}(\tau) .
 \label{MSCD_c}
 \end{eqnarray}
 Note that Eq.~(\ref{MSCD_c}) applies in the short  time scale regime, see discussion around Eq.~(\ref{MSCD_1}).

\section{Intramolecular probes}
\label{sec:Intramolecular-probes}
As an example of correlated pair of probes, we consider two tagged monomers belonging to a common polymer, see Fig.~\ref{fig2} (d). 

{\it Model---}.
Our calculation in this section is based on the Rouse model. The polymer consists of $N$ monomers connected through spring in series. The position of $n$-th monome $ {\vec r}(n,t)= (x(n,t), y(n,t), z(n,t))$ evolves according to the Rouse equation of motion, which takes the following form
\begin{eqnarray}
\gamma \frac{\partial {\vec r}(n,t)}{\partial t} = k \frac{\partial^2{\vec r}(n,t)}{\partial n^2} + {\vec \xi}(n,t)
\label{eq:Rouse}
\end{eqnarray}
where the friction coefficient $\gamma$ and the spring constant $k$ defines the monomeric time scale $\tau_0 = \gamma/k$. The last term ${\vec \xi}(n,t)$ is uncorrelated Gaussian white noise acting on $n$-th monomer with $\langle {\vec \xi}(n,t) \rangle = {\vec 0}$ and $\langle {\vec \xi}(n,t) {\vec \xi}(m,s) \rangle = 2 \gamma k_BT {\mathcal I} \delta (n-m) \delta (t-s)$, which represents the effect of thermal noise, with $k_BT$ being the thermal energy and  $3 \times 3$ identity matrix ${\mathcal I}$. Note that the above continuum approximation~(\ref{eq:Rouse}) is valid on length scale down to the monomer size $a$ with $k = 3 k_BT/a^2$.

In what follows, we neglect the chain end effect assuming large $N$. Then, starting from an initial configuration ${\vec r}(n,t_0)$ at time $t_0$, the solution is obtained as
\begin{eqnarray}
{\vec r}(n,t) &=& \frac{1}{\gamma} \int dn' \int_{t_0}^t dt'  \ G(n-n', t-t'){\vec \xi}(n',t')  \nonumber \\
&+&  \int dn' \ G(n-n', t-t_0){\vec r}(n',t_0)
\label{solutionR}
\end{eqnarray}
with the Green function
\begin{eqnarray}
G(n,t) = \sqrt{\frac{\gamma}{4 \pi k t}} \exp{\left(-\frac{\gamma}{4 k t} n^2 \right)}
\label{Gf}
\end{eqnarray}

{\it Mean square displacement---}.
Let us define the time correlation function of fluctuation $\delta x(n, t) = x(n,t) - \langle x(n,t)\rangle$ of a tagged monomer;
\begin{eqnarray}
C_0(t_0,t_1,t_2) \equiv \langle \delta x(n, t_1) \delta x(n, t_2) \rangle
\end{eqnarray} 
Note that this quantity generally depends on the initial time $t_0$ in addition to the two observation times $t_1$ and $t_2 ( \ge t_1)$. Using the solution~(\ref{solutionR}), it is calculated as
\begin{eqnarray}
&&C_0(t_0,t_1,t_2) = \frac{2 k_BT}{\gamma} \int_{t_0}^{t_1} dt_1' \ G(0,t_1+t_2-2t_1')  \nonumber \\
&=& \frac{k_BT}{k}\sqrt{\frac{1}{\pi \tau_0}} \left[ (t_1+t_2-2t_0)^{1/2} - (t_2-t_1)^{1/2} \right]
\end{eqnarray}
where we have used Eq.~(\ref{Gf}). From this, we obtain
\begin{eqnarray}
&&\langle( \delta x(n,t+\tau) -  \delta x(n,t))^2 \rangle  \nonumber \\
& =& C_0(t_0,t+\tau,t+\tau) + C_0(t_0,t,t) - 2C_0(t_0,t,t+\tau) \nonumber \\
&\rightarrow &  \frac{2 k_BT} {k}\sqrt{\frac{\tau}{\pi \tau_0}}  \qquad (t_0 \rightarrow - \infty), 
\end{eqnarray}
where in the final step, we focus on the stationary state by letting $t_0 \rightarrow -\infty$, in which the time translational invariance indicates $\langle x(n,t+\tau) \rangle =  \langle x(n,t) \rangle$, hence,  $ x(n,t+\tau) -  x(n,t)= \delta x(n,t+\tau) -  \delta x(n,t)$. This leads to the well-known subdiffusion scaling of MSD of each probe (with exponent $0.5$ for a Rouse model)~\cite{DoiEdwards,Panja2010,Saito2015};
\begin{eqnarray}
{\rm MSD}(\tau) &=& 3 \times \langle (x(n,t+\tau) - x(n,t))^2\rangle  \\
&=&\frac{2 a^2}{\sqrt{\pi}} \sqrt{\frac{\tau}{\tau_0}}, 
\label{MSD_Rouse}
\end{eqnarray}
where a factor $3$ reflects the dimensionality (the number of components) and the isotropy of the space, and we express the final expression in terms of monomer size $a$ and the corresponding time scale $\tau_0$. 

\begin{figure}[h]
	\centering
	\includegraphics[width=0.3\textwidth]{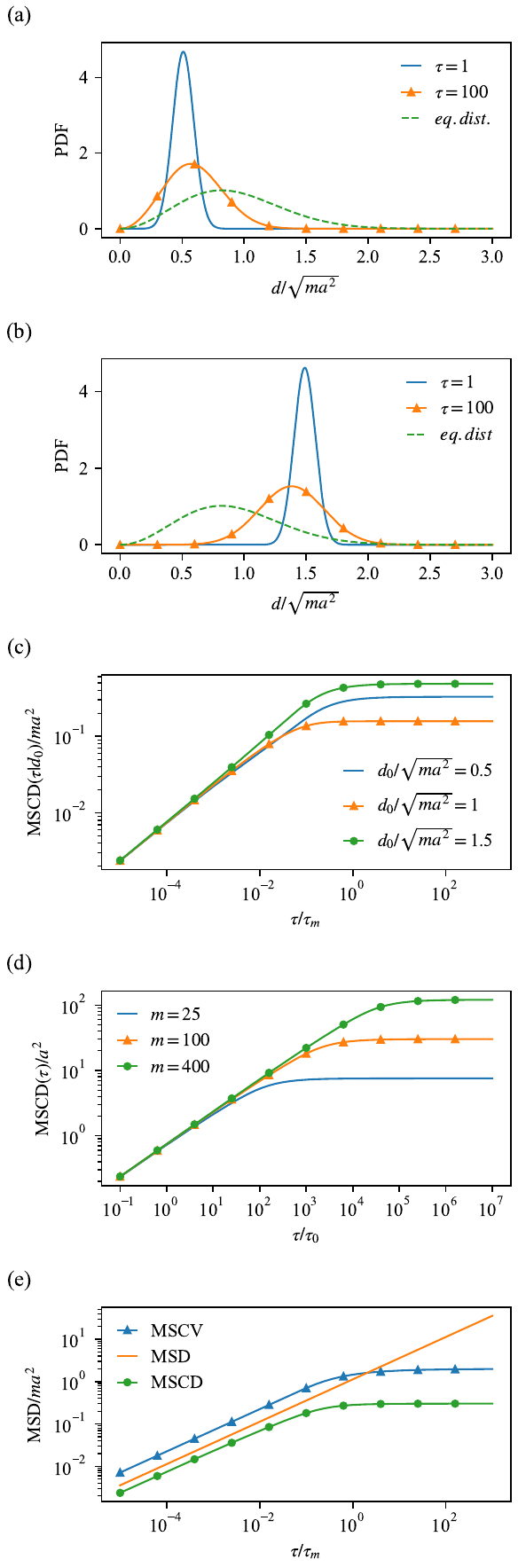}
	\caption{Two-point MSD analysis for a pair of intramolecular probes. (a) and (b) Time evolution of the probability density function $P_d(d,t | d_{0})$ of the inter-probe distance $d$ starting from (a) $d(0)=d_0=0.5 \sqrt{ma^2}$ and (b) $d(0)=d_0=1.5\sqrt{ma^2}$ . Dashed curves represent the equilibrium distribution. (c) Conditional $ {\rm MSCD}(\tau|d_0)$ as a function of scaled lag time $\tau/\tau_m$ for three different initial separation $d_0$. (d)  ${\rm MSCD}(\tau)$ as a function of lag time $\tau$ for three different inter-probe separation $m$. (e)$  {\rm MSCV}(\tau)$ and ${\rm MSCD}(\tau)$ as a function of scaled lag time $\tau/\tau_m$ compared to ${\rm MSD}(\tau)$.  }
	\label{fig5}
\end{figure}

{\it Correlation and relaxation functions of relative vector---}.
We label $n_1$-th and $n_2$-th monomers as two probes with their separation $m = |n_2-n_1|$, and assume the stationary state.
Central quantities in the following calculation are the time correlation function of $d_x(t)= x(n_1,t) - x(n_2,t)$;
\begin{eqnarray}
C_{\vec d}(\tau) \equiv \langle d_x(t_1+\tau) d_x(t_1) \rangle .
\label{C_d}
\end{eqnarray}
and the relaxation function
\begin{eqnarray}
{\tilde h}(\tau) \equiv \frac{\langle d_x(t_1+\tau)\rangle_{d_{0}}}{d_{x0}}
\label{h_tau}
\end{eqnarray}
where $\langle \cdots \rangle_{d_{0}}$ refers to the average over the sub-ensemble specified by the condition $d_x(t_1) = d_{x0}$.
In stationary state, both depend only on the separation of probes $m$ (aside from the neglected chain-end effect) and the time difference $\tau$.
In addition, these two quantities are related to each other via flucuation-dissipation theorem~\cite{DoiEdwards};
\begin{eqnarray}
\frac{m a^2}{3} {\tilde h}(\tau) = C_{\vec{d}}(\tau)
\label{FDT}
\end{eqnarray}
In Appendix~A, we calculate the relaxation function
\begin{eqnarray}
{\tilde h}(\tau) = \mathrm{erf}\left(\sqrt{\frac{\tau_m}{4 \tau}} \right) - \sqrt{\frac{ 4 \tau}{ \pi \tau_m}} \left[ 1- \exp{\left(- \frac{\tau_m}{4 \tau} \right)}\right], 
\label{h_relax}
\end{eqnarray}
where we introduce $\tau_m \equiv \tau_0 m^2$, which corresponds to the relaxation time of a subchain of size $m$.

\subsection{Relative vector analysis}
\label{sec:Intramolecular-probes_MSCV}
Summing up contributions from each component, ${\rm MSCV}(\tau) $ is calculated as
\begin{eqnarray}
{\rm MSCV}(\tau) &=& 3 \times \left[  (d_x(t+\tau) - d_x(t))^2\right] \nonumber \\
&=& 3 \times 2 \left[ C_{\vec d}(0) - C_{\vec d}(\tau)  \right] \nonumber \\
&=&2 ma^2  \left( 1 - {\tilde h}(\tau) \right) 
\label{MSCV_Rouse_1}
\end{eqnarray}
Equation~(\ref{MSCV_Rouse_1}) is consistent with the result reported in Ref.~\cite{Gabriele2022}.

Using limiting behavior of ${\tilde h}(\tau)$, we find
\begin{eqnarray}
{\rm MSCV}(\tau) \simeq \left\{
 \begin{array}{ll}
 \frac{4 a^2}{\sqrt{\pi}}  \sqrt{\frac{\tau}{ \tau_0}} & ( \tau \ll \tau_m) \\
2 ma^2 & (\tau \gg \tau_m)
 \end{array}
 \right.
\label{MSCV_Rouse_2}
\end{eqnarray}
Comparing Eqs.~(\ref{MSD_Rouse}) and~(\ref{MSCV_Rouse_2}), one find ${\rm MSCV}(\tau) = 2 {\rm MSD} (\tau)$ in short time-scale regime ($\tau \ll \tau_m$), the same relation to the case of independent probes (Sec.~\ref{sec:Independent-probes_MSCV}), see Fig.~\ref{fig5} (e).

\subsection{Distance analysis}
\label{sec:Intramolecular-probes_MSCD}
We will follow the steps illustrated in Sec.~\ref{sec:Independent-probes_MSCD} for the case of independent probes. We first need to find $P_{{\vec d}}({\vec d},\tau | {\vec d}_{0})$; the probability density function of the relative vector ${\vec d}$ at $t=t_1+\tau$ given the condition $|{\vec d}(t=t_1)|=  d_0$ in stationary state.
Let $t_0 \rightarrow - \infty$ in Eq.~(\ref{solutionR}), the system is thus in equilibrium at $t=t_1$. From this equilibrium ensemble, we pick up a sub-ensemble filtered by the condition ${\vec d}(t_1) = d_0/\sqrt{3}(1,1, 1)$. From here on, we measure the time as a function of $\tau=t-t_1$ since we are interested in the subsequent time evolution after the initial condition at $t=t_1$.
The probability density function takes the form
\begin{eqnarray}
P_{{\vec d}}({\vec d},\tau | {\vec d}_{0}) = \frac{1}{[2 \pi g(\tau)]^{3/2}} \exp{\left( - \frac{ \sum_{\alpha = x,y,z} [d_{\alpha} - \langle d_{\alpha}(\tau) \rangle_{d_0}]^2}{ 2 g(\tau)}\right)} \nonumber \\
\end{eqnarray}
where $ \langle \cdots \rangle_{d_0}$ refers to averaging over the sub-ensebmle, and $g(\tau) \equiv \langle \{ d_{\alpha}(\tau) - \langle  d_{\alpha}(\tau) \rangle_{d_0} \}^2 \rangle$.
Note that the ``initial" separation at $\tau=0$ shows only up through the first moment, i.e., the average time evolution $\langle d_{\alpha}(\tau) \rangle_{d_0}$  of each component ($\alpha = x, y, z$). In our ``initial" condition, all the components follows the same average evolution, which is described using the relaxation function, see Eq.~(\ref{h_tau});
\begin{eqnarray}
\langle d_{\alpha}(\tau) \rangle_{d_0} =   \frac{d_0}{\sqrt{3}} {\tilde h}(\tau)  \qquad (\tau \ge 0, \ \alpha=x,y,z)
\end{eqnarray}

 The variance $g(\tau) \equiv \langle \{ d_{\alpha}(\tau) - \langle  d_{\alpha}(\tau) \rangle_{d_0} \}^2 \rangle$ is calculated as 
\begin{eqnarray}
g(\tau) &=& \langle \left[ \{d_{\alpha}(\tau) - d_{\alpha}(0) \} + \{d_{\alpha}(0) -   \langle  d_{\alpha}(\tau) \rangle \} \right]^2 \rangle \nonumber \\
&=& \langle \{d_{\alpha}(\tau) - d_{\alpha}(0) \}^2\rangle - \{d_{\alpha}(0) -   \langle  d_{\alpha}(\tau) \rangle \}^2  \nonumber \\
&=& 2 (C_{{\vec d}}(0) - C_{{\vec d}}(\tau)) - \frac{ma^2}{3} (1 - {\tilde h}(\tau))^2 \nonumber \\
&=&  \frac{m a^2}{3}(1- {\tilde h}(\tau)^2).
\end{eqnarray}
where we have used Eq.~(\ref{FDT}) in the final step.
Note that, to obtain the second term in third line, we have carried out the following averaging
\begin{eqnarray}
\int_0^{\infty} \mathrm{d}d_0 \ d_0^2 \ p(d_0) = ma^2
\end{eqnarray}
with the equilibrium distribution of $d_0$;
\begin{eqnarray}
p(d_0) = 4 \pi \left( \frac{3}{2 \pi m a^2}\right)^{3/2} d_0^2 \ \exp{\left( - \frac{3 d_0^2}{2 ma^2}\right)}
\label{p_d0}
\end{eqnarray}

After variable transformation from ${\vec d}$ to $d = \sqrt{|{\vec d}|^2}$, we obtain the probability density function of $d$ conditioned on $d(0) = d_0$;
\begin{eqnarray}
&&P_d(d,\tau|d_0) = \frac{d}{d_0 {\tilde h}(\tau)\sqrt{2 \pi g(\tau)}}  \nonumber \\
&&  \times \left[\exp{\left(- \frac{(d- d_0 {\tilde h}(\tau))^2}{2 g(\tau)} \right)}  - \exp{\left(- \frac{(d+ d_0 {\tilde h}(\tau))^2}{2 g(\tau)} \right)} \right] \nonumber \\
\label{P_d_intra}
\end{eqnarray}
Figure~\ref{fig5} (a) or (b) shows the time evolution of $P_d(d,\tau|d_0)$ starting from a shrunk ($d_0/\sqrt{ma^2} < 1$) or swollen ($d_0/\sqrt{ma^2} > 1$) conformation, respectively. Unlike for the case of independent probes, the distributions in both cases approach towards the equilibrium distribution given by Eq.~(\ref{p_d0}) after $\tau \simeq \tau_m$. From this, we obtain the MSCD conditioned on $d(0) = d_0$, see Fig.~\ref{fig5} (c);
 \begin{eqnarray}
&& {\rm MSCD} (\tau| d_0)  \equiv \int_0^{\infty} \mathrm{d}d \ (d - d_0)^2 P_d(d,\tau | d_{0})  \nonumber \\
&=& 3 g(\tau) + d_0^2(1+ {\tilde h}(\tau)^2) -  d_0 \sqrt{\frac{8 g(\tau)}{\pi}} \exp{\left(-\frac{(d_0 {\tilde h}(\tau))^2}{2g(\tau)}\right) }  \nonumber \\
&&- 2\left[\frac{g(\tau) + (d_0 {\tilde h}(\tau))^2}{{\tilde h}(\tau)} \right] \mathrm{erf}\left( \frac{d_0{\tilde h}(\tau)}{\sqrt{2 g(\tau)}}\right)
\label{MSCD_d0_intra}
\end{eqnarray}
Comparing Eqs.~(\ref{P_d_intra}) and~(\ref{MSCD_d0_intra}), respectively, with Eqs.~(\ref{P_d_indep}) and~(\ref{MSCD_d0_indep}), it is clear how the connectivity effect comes in the statistical time evolution of inter-probe distance.

Experimentally, as discussed in Sec.~\ref{sec:Independent-probes_MSCD}, more easily accessible quantity is
the time averaged MSCD, see Eq.~(\ref{MSCD_ta}), obtained by averaging over the equilibrium distribution of $d_0$ given by Eq.~(\ref{p_d0}).
 After performing the averaging operation, we find
\begin{eqnarray}
&&\frac{{\rm MSCD} (\tau)}{m a^2} \nonumber \\
&=& 2 - \frac{4}{\pi}\sqrt{1-{\tilde h}(\tau)^2}\left[ 1 + \left( \omega({\tilde h}) + \frac{1}{3 \omega ({\tilde h})} \right) \arctan{(\omega ({\tilde h}))}\right]  \nonumber \\
\label{MSCD_Rouse}
\end{eqnarray}
where we have defined $\omega(x) = \sqrt{x^2/(1-x^2)}$. Using limiting behaviors of ${\tilde h}(\tau)$, we find
\begin{eqnarray}
{\rm MSCD} (\tau)  \simeq \left\{
 \begin{array}{ll}
 \frac{4 a^2}{3 \sqrt{\pi}}  \sqrt{\frac{\tau}{ \tau_0}} & ( \tau \ll \tau_m) \\
ma^2 \left(2 - \frac{16}{3 \pi}  \right) & (\tau \gg \tau_m)
 \end{array}
 \right.
\label{MSCD_Rouse_2}
\end{eqnarray}
Comparing Eq.~(\ref{MSCD_Rouse_2}) with Eq.~(\ref{MSD_Rouse}), we find ${\rm MSCD}(\tau) = 2/3 {\rm MSD} (\tau)$ in short time-scale regime ($\tau \ll \tau_m$), the same relation to the case of independent probes (Sec.~\ref{sec:Independent-probes_MSCD}), see Fig.~\ref{fig5} (e).

\section{Discussions}
\label{sec:DIscussions}
\subsection{General property of MSCV and MSCD}
\label{GP_MSCD}
\begin{figure}[h]
	\centering
	\includegraphics[width=0.3\textwidth]{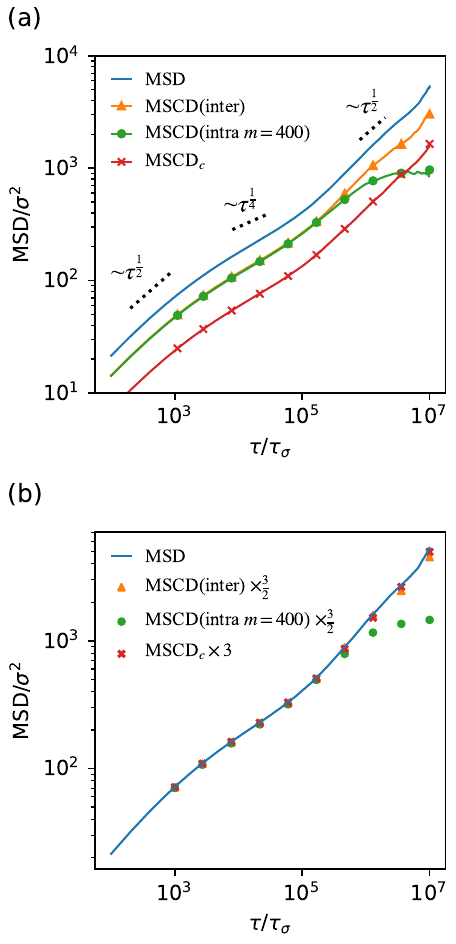}
	\caption{Distance analysis performed for tagged monomer probes in entangled polymer solutions. (a) Shown are the following quantities as a function of lag time $\tau$; ${\rm MSD}$ of a tagged monomer,  ${\rm MSCD}$(inter) from two independent probes on different polymers, ${\rm MSCD}$(intra) from two intramolecular probes, ${\rm MSCD_c}$ from a single probe and the center of mass. Selection of tagged monomer(s) along each chain (with length $N=600$) is $n=100$, or $(n_1, n_2)=(100, 500)$ for the intramolecular pair. (b) All the quantities shown in (a) collapse according to formulas~(\ref{MSCD_MSD}) and~(\ref{MSCD_c}). }
	\label{fig6}
			\vspace{0.2 cm}
\end{figure}
In Sec.~\ref{sec:relation_MSD}, we have claimed the formula~(\ref{MSCD_MSD}) for the relation between MSD, MSCV and MSCD.
So far, we have verified it for two independently moving probes in Sec.~\ref{sec:Independent-probes}, see also Eqs.~(\ref{MSCV_c}) and~(\ref{MSCD_c}) for the analysis in the center-of-mass frame (Sec.~\ref{sec:CM_frame}), and for the intramolecular pair of probes within a polymer in short-time scale ($\tau \ll \tau_m$) in Sec.~\ref{sec:Intramolecular-probes_MSCD}.
Although our analyses in these preceding sections are based on simple models, i.e., Brownian motion model in Sec.~\ref{sec:Independent-probes} and Rouse model in Sec.~\ref{sec:Intramolecular-probes},  we now discuss that the relation~(\ref{MSCD_MSD}) is quite generally expected even for more complex situations, i.e., probe particles performing anomalous diffusion (see, for instance, Fig.~\ref{fig2} (b)) and probes belonging to more general polymers, etc., as a consequence of defining properties of MSCV and MSCD.
For the latter, the discussion allows us to provide the precise definition of the short time scale mentioned in Sec.~\ref{sec:relation_MSD}, and clarify the underlying physics behind the formula~(\ref{MSCD_MSD}).

We begin by repeating a simple argument in Sec.~\ref{sec:relation_MSD} for the relation between MSD and MSCV.
Since $\Delta {\vec d}(\tau) = \Delta {\vec r}_1(\tau) - \Delta {\vec r}_2(\tau)$, one can transform Eq.~(\ref{MSCR1}) as
\begin{eqnarray}
{\rm MSCV}(\tau)& =& \langle | \Delta {\vec r}_1( \tau) |^2\rangle + \langle | \Delta {\vec r}_2( \tau) |^2\rangle  - 2 \langle \Delta {\vec r}_1(\tau) \cdot \Delta {\vec r}_2(\tau) \rangle \nonumber \\
\label{MSCV_eq}
\end{eqnarray}
Then, assuming that two probes are independent; $ \langle \Delta {\vec r}_1(\tau) \cdot \Delta {\vec r}_2(\tau) \rangle =  \langle \Delta {\vec r}_1(\tau)\rangle \cdot  \langle\Delta {\vec r}_2(\tau) \rangle  = 0$,
and have the same diffusivity, one finds ${\rm MSCV}(\tau) = 2 {\rm MSD} (\tau)$.

For the relation between MSD and MSCD, we start from a relation ${\vec d}(t_0 + \tau) = {\vec d}(t_0) + \Delta {\vec r}_1(\tau) - \Delta {\vec r}_2(\tau)$, and introduce a unit vector ${\hat u}$ associated with the relative vector at $t=t_0$, i.e., ${\vec d}(t_0) = d_0 {\hat u}$, see also Refs.~\cite{Ponisch2018, Yesbolatova2022} for a related discussion. Then, the distance at time $t= t_0 + \tau$ is expressed as
\begin{eqnarray}
d(t_0+\tau) &= &d_0 \sqrt{\left|  {\hat u} + \frac{\Delta {\vec r}_1(\tau) - \Delta {\vec r}_2(\tau)}{d_0}\right|^2} \nonumber \\
&\simeq& \left\{
 \begin{array}{ll}
d_0 + (\Delta {\vec r}_1(\tau) - \Delta {\vec r}_2(\tau)) \cdot {\hat u} & ( \kappa \ll 1)  \\
 |\Delta {\vec r}_1(\tau) - \Delta {\vec r}_2(\tau)| & ( \kappa \gg 1)
 \end{array}
 \label{d_app}
 \right.
 \nonumber \\
\end{eqnarray}
where we naturally introduce a dimensionless quantity $\kappa \equiv |\Delta {\vec r}_1(\tau) - \Delta {\vec r}_2(\tau)|/d_0$, which enables us to discriminate the short or long length and time scales for a given $d_0$. Equation~(\ref{d_app}) has a clear physical interpretation; in short time scale ($\kappa \ll 1$), relevant in the change in $d$ is the projected motion along ${\vec d}(t_0)$, while $d_0$ is completely irrelevant in the long time limit  ($\kappa \gg 1$) . For convenience, we set $ {\hat u} = (1,0,0)$ without loss of generality.  One thus finds, for a pair of independently moving probes,
\begin{eqnarray}
&&{\rm MSCD}(\tau|d_0) \nonumber \\
& \simeq&
\left\{
 \begin{array}{ll}
\langle (\Delta x_1(\tau) - \Delta x_2(\tau))^2 \rangle  = \frac{2}{3} \ {\rm MSD} (\tau) & ( \kappa \ll 1)  \\
 \langle |\Delta {\vec r}_1(\tau) - \Delta {\vec r}_2(\tau)|^2 \rangle  = 2  \ {\rm MSD} (\tau) & ( \kappa \gg 1) 
 \end{array}
 \right. 
 \nonumber \\
\end{eqnarray}
This generalizes Eq.~(\ref{MSCD_d}); in long-time regime ($\kappa \gg 1$), ${\rm MSCD}(\tau|d_0)$ is the same as ${\rm MSCV}(\tau)$ and equal to twice ${\rm MSD}(\tau)$. On the other hand, a reduction factor $1/3$, or more generally, 1/(space dimension), appears in short-time regime ($\kappa \ll 1$), since we only count one component, i.e., along ${\vec d}(t_0)$ in the calculation. As we have seen in Sec.~\ref{sec:Independent-probes_MSCD}, this short-time scale behavior dominates after averaging over $d_0$ for a pair of independent probes in stationary state.  Although the pair of intramolecular probes analyzed in Sec.~\ref{sec:Intramolecular-probes_MSCD} belong to the same chain, their behaviors are essentially independent in the short-time scale $\tau \ll \tau_m$, as Eq.~(\ref{Gf}) implies. Hence, in both cases, one expect the relation~(\ref{MSCD_MSD}).
On the other hand, a pair of intramolecular probes is no longer independent of each other at $\tau > \tau_m$, hence, ${\rm MSCD}(\tau)$ as well as ${\rm MSCV}(\tau)$  eventually approach a $m$-dependent constant value.

In Fig.~\ref{fig6}, we show an example of MSCD analysis applied to the trajectory data obtained from molecular dynamic simulation of entangled polymer solutions. Here we employ a rather standard model of polymer solution with the chain length $N=600$ and the monomer volume fraction $\phi = 0.1$ (see Appendix B for more details). As shown, MSD of tagged monomer exhibits rich anomalous behaviors, i.e., early $\tau^{1/2}$, subsequent slowing-down, a signature of the reptation dynamics (${\rm MSD} \sim \tau^{1/4}$) that is followed by later ${\rm MSD} \sim \tau^{1/2}$ scaling~\cite{DoiEdwards, de_Gennes_scaling}. In addition to MSD, we have calculated ${\rm MSCD}$ using probes on different polymers (see Fig.~\ref{fig2} (b)) and on the same polymer (see Fig.~\ref{fig2} (d)). We have also calculated ${\rm MSCD_c}$ from the distance analysis between a probe and the center of mass of the system  (see Fig.~\ref{fig2} (c)). It is remarkable that all these quantities can be mapped to MSD quite accurately according to the formulas~(\ref{MSCD_MSD}) and~(\ref{MSCD_c}).

\subsection{Non-Gaussian parameter}
Recently, there has been growing interest on the motion of probes in heterogeneous environment, where the statistics of probe displacement often exhibits non-Gaussian distribution~\cite{Wang2009, Chubynsky2014, Chechkin2017}. This may be also relevant to the motion of chromatin loci in nucleus.  It is therefore tempting to develop a method to quantify the non-Gaussianity within the scheme of two-point MSD.
Here, we outline such a method through the distance analysis, which can be applied even when the rotational as well as the translational motion of the frame can not be neglected.

Since the short-time scale regime $\kappa \ll 1$ would eventually dominate upon averaging over $d_0$, let us focus our attention on such a regime.  From Eq.~(\ref{d_app}) we obtain the fourth moment of the distance change
\begin{eqnarray}
\langle \Delta d (\tau)^{4} \rangle  & \simeq& \langle (\Delta x_1(\tau) - \Delta x_2(\tau))^{4} \rangle \nonumber \\
&=&  \langle \Delta x_1(\tau)^4 \rangle +  6 \langle \Delta x_1(\tau)^2\rangle  \langle \Delta x_2(\tau)^2 \rangle \nonumber \\
&& +  \langle \Delta x_2(\tau)^4 \rangle 
\end{eqnarray}
where we have used the independence of $\Delta x_1(\tau)$ and $\Delta x_2(\tau)$ and their third moments vanish due to symmetry.
If the displacement follows the Gaussian distribution, the relation $\langle \Delta x_i(\tau)^4 \rangle = 3 \langle \Delta x_i(\tau)^2 \rangle^2 $ (for $ i = 1, 2$) simplifies the above equation to
\begin{eqnarray}
\langle \Delta d (\tau)^{4} \rangle   \xrightarrow[{\rm Gaussian}]{}&& 3 \left[ \langle \Delta x_1(\tau)^2 \rangle + \langle \Delta x_2(\tau)^2 \rangle \right]^2  \nonumber \\
&=& \frac{4}{3} [{\rm MSD}(\tau)]^2
\end{eqnarray}
Combining with Eq.~(\ref{MSCD_MSD}), this leads us to define the distance-analysis-based non-Gaussian parameter
\begin{eqnarray}
\Delta^{{\rm NG}}_d(\tau) &\equiv& \frac{\langle \Delta d (\tau)^4 \rangle }{ 3[{\rm MSCD}(\tau)]^2} -1 
\label{NGP_d}
\end{eqnarray}

\begin{figure}[h]
	\centering
	\includegraphics[width=0.3\textwidth]{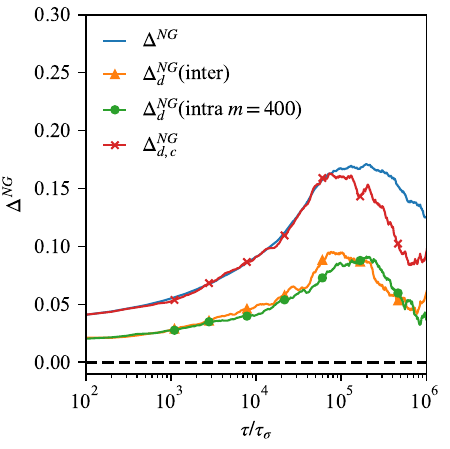}
	\caption{Non-Gaussian parameters calculated for the molecular dynamics simulation data of entangled polymer solutions. Shown are non-Gaussian parameters calculated from Eq.~(\ref{NGP_d}) and Eq.~(\ref{NGP_single}). For the former, we employed (i) independent probe pair, (ii) intramolecular probe pair, (iii) a single probe and the center of mass to define the distance. Selection of tagged monomer(s) along each chain (with length $N=600$) is $n=100$, or $(n_1, n_2)=(100, 500)$ for the intramolecular pair.}
	\label{fig8}
			\vspace{0.2 cm}
\end{figure}

We applied the analysis to the entangled linear polymer solution, in which the entanglement effect is known to cause non-Gaussian dynamics~\cite{Shen2022}.  Employing the same data used in Fig.~\ref{fig6} (chain length $N=600$, monomer volume fraction $\phi=0.1$), we calculated the distance-based non-Gaussian parameter from (i) independent probe pair, (ii) intramolecular probe pair, (iii) a single probe and the center of mass. The results are shown in Fig.~\ref{fig8}.
Compared to a standard non-Gaussian parameter in single particle tracking analysis~\footnote{This definition comes from the following observation. If the probe displacement is isotropic in space and follows Gaussian statistics, the following calculation holds; $\langle |\Delta {\vec r}(\tau)|^4 \rangle = \langle [\Delta x(\tau)^2 + \Delta y (\tau)^2 + \Delta z(\tau)^2]^2 \rangle = 3 \langle \Delta x(\tau)^4\rangle + 6 \langle \Delta x(\tau)^2 \rangle^2 = 15 \langle \Delta x(\tau)^2 \rangle^2 = 5 [{\rm MSD}(\tau)]^2/3$.  }
\begin{eqnarray}
\Delta^{{\rm NG}} (\tau) \equiv \frac{3 \langle |\Delta {\vec r}(\tau)|^4 \rangle }{5 [{\rm MSD}(\tau)]^2} -1
\label{NGP_single}
\end{eqnarray}
all the distance-based non-Gaussian parameters correctly captures the growing non-Gaussianity with the time scale and the peak position around the Rouse time $\sim \tau_0 N^2$.

 In Ref.~\cite{Yesbolatova2022}, authors observed the motion of chromatin loci in {\it C. elegans} embryo, and found a deviation from Rouse model behavior after 8 cell stage. Concomitantly, cells start to organize various characteristic intranuclear structures such as nucleolus, heterochromatin foci, etc~\cite{Arai2017}. It would be interesting to see whether the qualitative change in loci dynamics may be associated with the possible non-Gaussian displacement statistics due to the nascent inhomogeneous nuclear environment.

\subsection{Motional correlation}
\label{DispCorr}
In the discussion of two-point MSD, one of the central quantities is the correlation in the motion of probes. Here, it is important to notice that there are two types of correlations, i.e., either extrinsic arising from substrate motion or intrinsic correlation between probes. In this paper, we have been mostly concerned with the correlation of extrinsic origin. Indeed, a basic idea of two-point MSD analysis is filtering out such a correlation arising from the substrate motion. Specifically, MSCD has the property ${\rm MSCD}(\tau) = {\rm MSCD}^{(o)}(\tau)$, thus, can be calculated from $\Delta {\vec r}_i^{(o)}(\tau)$ for $i=1,2$ (see Fig.~\ref{fig3}), with which one can obtain MSD using Eq. (\ref{MSCD_MSD}) as long as there is no intrinsic correlation in their dynamics.
As a different, but related methodology, Oliveira et al. has recently proposed a framework that, through the analysis of correlated motion of probes, provides a way to estimate the substrate motion, giving an access to the intrinsic probe mobility~\cite{Oliveira2021}.

In the remainder of this subsection, we will focus on the opposite situation, where the substrate motion is negligible, i.e.,  $\Delta {\vec r}^{(o)} (\tau) = \Delta {\vec r} (\tau)$, thus, not only MSCD but also MSD and MSCV can be calculated from $\Delta {\vec r}^{(o)}(\tau)$, but there is intrinsic correlation in the dynamics of probe pairs, invalidating the relation
(\ref{MSCD_MSD}).
In turn, this indicates that two-point MSD analysis can be combined with the standard MSD analysis to quantify the degree of  intrinsic correlation between motion of two probes. 


We illustrate the idea using the MSCV as a two-point MSD output.
Assuming equal mobility for two probes,  Eq~(\ref{MSCV_eq}) leads to the expression of displacement correlation between two probes in the time-scale $\tau$ in terms of ${\rm MSD}(\tau)$ and ${\rm MSCV}(\tau)$;
\begin{eqnarray}
\langle \Delta {\vec r}_1(\tau) \cdot \Delta {\vec r}_2(\tau) \rangle = {\rm MSD}(\tau) - \frac{{\rm MSCV}(\tau)}{2}
\end{eqnarray}

\begin{figure}[h]
	\centering
	\includegraphics[width=0.3\textwidth]{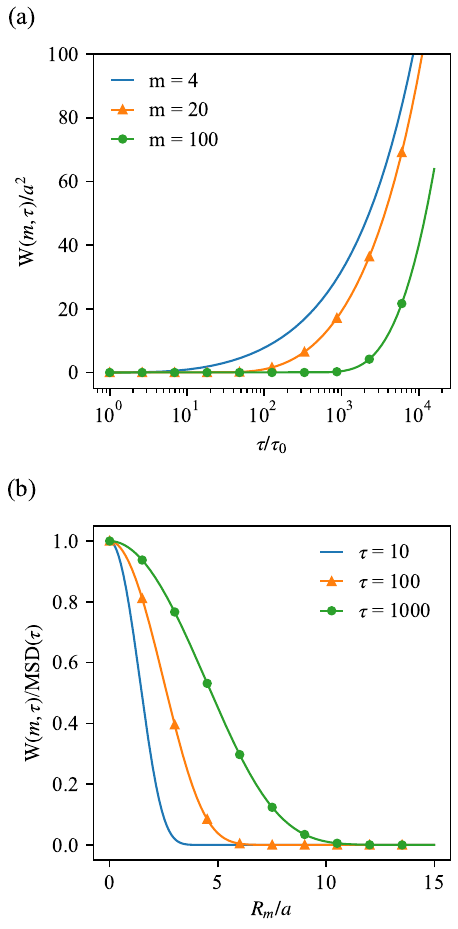}
	\caption{ Displacement correlation $W(m, \tau)$ for a pair of tagged monomers $m$ apart along backbone in Rouse polymer. (a) $W(m, \tau)$ as a function of time-scale $\tau$ for fixed values of $m=4, 20, 100$. (b) Normalized displacement correlation $W(m, \tau) /{\rm MSD}(\tau)$ as a function of characteristic spatial distance $R_m=a m^{1/2}$ between two tagged monomers for fixed values of $\tau=10, 100, 1000$.}
	\label{fig7}
			\vspace{0.2 cm}
\end{figure}

We now apply the scheme to the intramolecular probe pair within a single Rouse chain.
Let us define $W(m,\tau) \equiv \langle \Delta {\vec r}(n_1, \tau) \cdot \Delta {\vec r}(n_2,\tau) \rangle$ as the correlation between displacements in the time-scale $\tau$ of two tagged monomers separated $m=|n_1-n_2|$ along the chain. Using Eqs.~(\ref{MSD_Rouse}) and~(\ref{MSCV_Rouse_1}), we obtain
\begin{eqnarray}
W(m,\tau) = ma^2 \left[  \sqrt{\frac{4 \tau}{\pi \tau_m}} \exp{\left( - \frac{\tau_m}{4 \tau}\right)} -{\rm erfc} \left( \sqrt{\frac{\tau_m}{4 \tau}}\right)  \right] \nonumber \\
\end{eqnarray}
where ${\rm erfc}(z) = 1-{\rm erf}(z)$ is complementary error function.
In Fig.~(\ref{fig7}) (a), we plot $W(m,\tau)$ as a function of $\tau$ for three different inter-probe separations. The correlation between the probes' displacement is almost zero in the short time-scale in all cases, but starts to rise at the characteristic time-scale $\tau \simeq \tau_m$. Two probes thus do not feel each other in the time-scale $\tau < \tau_m$, but move together in longer time-scale, in which $W(m,\tau) \sim {\rm MSD}(\tau)$.
One can also look at $W(m,\tau)$ as a function of $m$ for a given $\tau$, or in the same way as a function of $R_m$ that represents the characteristic spatial distance between two tagged monomers $m$-apart along the chain. With $W(0,\tau) = {\rm MSD}(\tau)$, the normalized quantity $W(m,\tau)/{\rm MSD}(\tau)$ represents the degree of motional correlation in the time-scale $\tau$ for two tagged monomers with spatial distance $R_m = am^{1/2}$ apart.  In Fig.~(\ref{fig7}) (b), we plot $W(m,\tau)/{\rm MSD}(\tau)$ as a function of $R_m$ for three different time-scales. One can identify the characteristic ``domain", whose size grows with time-scale as $\sim a \tau^{1/4}$.

In Ref.~\cite{Nozaki2023}, a similar analysis based on two-point MSD measurement has been performed in an attempt to identify the domain size of chromatin.
It should be of great interest to see how the correlation seen in real chromatin would be compared to the result for Rouse polymer, and how the deviation, if any, can be connected to the spatial organization of chromatin domain in nucleus.

\section{Summary}
The growing need for two-point MSD has been seen in several recent experiments, in particular those that track chromatin loci in living cell nucleus~\cite{Marshall1997, Mine-Hattab2012, Ochiai2015, Arai2017, Khanna2019, Yesbolatova2022, Gabriele2022, Bruckner2023, Nozaki2023, Minami2024}. Yet, the term two-point MSD has so far been used depending on context with nuances, lacking its precise definition.  
In this note, we have summarized the idea of two-point MSD, the definition of two variants ${\rm MSCV}$ and ${\rm MSCD}$ and their basic properties for a pair of independent probes as well as intramolecular probes. 
As is clear from their definitions, ${\rm MSCD}$ is more suitable to quantify the intrinsic mobility of the probe when the rotational motion of the substrate comes in. Otherwise, ${\rm MSCV}$ can be used alike, see the demonstration in Sec.~\ref{sec:Quantities}.  
We note that a apparently similar set-up has been conventionally employed in micro-rheology experiments but with a different motivation. Indeed, while the two-point micro-rheology is mainly motivated to eliminate the influence of local inhomogeneities in complex media~\cite{Crocker2000}, the two-point MSD discussed here aims at providing a way to quantify the intrinsic mobility of probes by eliminating the influence of the extrinsic motion.

Simple reasoning in Sect.~\ref{GP_MSCD} shows that the relation~(\ref{MSCD_MSD}) among ${\rm MSCV}$, ${\rm MSCD}$ and ${\rm MSD}$ generally holds for a pair of independent probes moving in free space in stationary state (or in systems with stationary increment).
For pairs that are not independent, on the other hand,  the relation~(\ref{MSCD_MSD}) breaks down. In that case, one can exploit this fact by combining two-point MSD measurement with ${\rm MSD}$, which enables us to extract and quantify the degree of motional correlation between those probe pairs (see Sec.~\ref{DispCorr}).

As necessary conditions for the relation~(\ref{MSCD_MSD}), in addition to the ``independence", we put the "free space" and the "stationarity" as our analysis does not take into account the effect of the geometry of the confining space, and we also need the average over the stationary distribution of $d_0$ to obtain the (time averaged) ${\rm MSCD}$. It would be interesting to see how the ${\rm MSCD}$ behaves in nonstationary, e.g., aging systems. The effect of confinement will show up when the typical displacement becomes comparable to the confinement size. 

We have also proposed a way to probe non-Gaussianity in displacement statistics based on the distance analysis. We hope that the method proves to be useful in characterizing dynamics of systems embedded in moving frame as in the case of chromatin in cell nucleus.

\section*{Acknowledgements}
This work was supported by JSPS KAKENHI (Grant Nos. JP23H00369, JP23H04290 and JP24K00602) and AGU Future Eagle Project. We thank A. Kimura for discussion and motivating us to summarize this note, and C. Lim for insightful comments on our note.

\section*{Appendix}
\subsection{Relaxation function}
\label{App1}

\begin{figure}[h]
	\centering
	\includegraphics[width=0.3\textwidth]{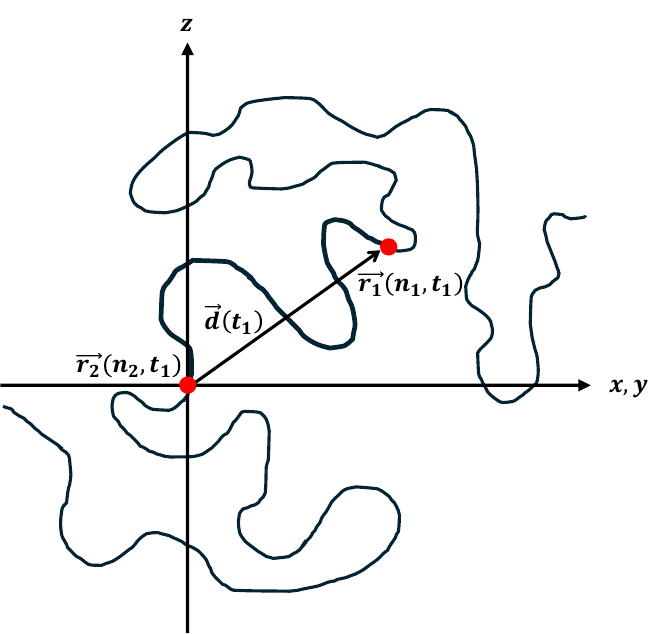}
	\caption{Schematic representation of a polymer configuration for the ``initial condition" at $t=t_1$.}
	\label{fig9}
			\vspace{0.2 cm}
\end{figure}

Here we calculate the relaxation function of the relative vector ${\vec d}(t)={\vec r}(n_1,t) - {\vec r}(n_2,t)$ in equilibrium state. From the equilibrium ensemble (the canonical distribution of ${\vec d}$ with $\langle {\vec d} \rangle = {\vec 0}$), we pick up sub-ensemble of configurations that satisfy  $|{\vec d(t_1)}| = d_0$, which is created by the spontaneous fluctuations. The relaxation function describes the subsequent average evolution of ${\vec d(t)}$ ($t > t_1$) in this sub-ensemble towards the equilibrium value ${\vec d} = {\vec 0}$.

Without loss of generality, we assume $n_1 > n_2$ and set ${\vec r}(n_1,t_1) = (d_0/\sqrt{3})(1,1,1)$ and ${\vec r}(n_2,t_1) = (0,0,0)$. This implies the ``initial condition" at time $t_1$ for each Cartesian component ($\alpha = x, y, z$)
\begin{eqnarray}
\langle r_{\alpha}(n,t_1) \rangle_{d_0}  
& =&
\left\{
 \begin{array}{ll}
\frac{d_0}{3} & ( n_1 < n)  \\
\frac{n-n_2}{n_1-n_2}\frac{d_0}{3}   & ( n_2 < n \le n_1)  \\
0 & ( n \le n_2)  \\
 \end{array}
 \right. 
 \label{IC_d_0}
\end{eqnarray}
where the subscript $d_0$ indicates the average over the sub-ensemble (Fig.~\ref{fig9}).

From Eq.~(\ref{solutionR}), the average position of $n$-th monomer evolves according to
\begin{eqnarray}
\langle r_{\alpha}(n,t_1+\tau) \rangle_{d_0} = \int dn' \  G(n-n',\tau) \langle r_{\alpha}(n',t_1) \rangle_{d_0}
\end{eqnarray}
Using Eq.~(\ref{Gf}) for Green function and the form~(\ref{IC_d_0}) for the ``initial condition", we find the average position of two probe monomers as
\begin{eqnarray}
\langle r_{\alpha}(n_1,t_1+\tau) \rangle_{d_0} =&& \frac{d_0}{2 \sqrt{3}}  \left[ \mathrm{erf}{\left( \sqrt{\frac{\tau_m}{4 \tau}}\right)} + 1\right] \nonumber \\
&&+ \frac{d_0}{2 \sqrt{3 \pi}} \sqrt{\frac{4 \tau}{\tau_m}} \left( e^{-\frac{\tau_m}{4 \tau}} -1\right) \nonumber \\
\label{r_1_ave}
\end{eqnarray}
\begin{eqnarray}
\langle r_{\alpha}(n_2,t_1+\tau) \rangle_{d_0} =&& \frac{d_0}{2 \sqrt{3}}  \left[ 1- \mathrm{erf}{\left( \sqrt{\frac{\tau_m}{4 \tau}}\right)} \right] \nonumber \\
&&+ \frac{d_0}{2 \sqrt{3 \pi}} \sqrt{\frac{4 \tau}{\tau_m}} \left( 1- e^{-\frac{\tau_m}{4 \tau}} \right) \nonumber \\
\label{r_2_ave}
\end{eqnarray}
where $\tau_m = \tau_0 m^2$ with $m = n_1-n_2$.
Equations~(\ref{r_1_ave})and~(\ref{r_2_ave}) together with the definition of relaxation function in Eq.~(\ref{h_tau}) leads to Eq.~(\ref{h_relax}).

\subsection{Simulation of probe dynamics in moving substrate}
\label{App2-1}
We consider the motion of probe particle undergoing fractional Brownian motion (fBm). Denoting the probe trajectory as ${\vec x}(t)$, fBm is characterized by the stationary increment $\Delta {\vec x}(\tau) = {\vec x}(t_0+\tau) - {\vec x}(t_0)$, whose auto-correlation is given by
\begin{eqnarray}
\langle \Delta x(\tau) \Delta x(\tau') \rangle = K_{\alpha}(\tau^{\alpha} + \tau'^{\alpha} - |\tau-\tau'|^{\alpha})
\end{eqnarray}
for each Cartesian component with $\alpha$ being the twice of the so-called Hurst exponent, the time and the length to be measured in appropriate units. If we assume fBm as a model of a tagged monomer in polymer (see Sec.~\ref{sec:Intramolecular-probes}), we may identify monomeric time and length scales ($\tau_0$ and $a$, respectively) as these units. 

To describe the translational motion of the substrate, let ${\vec X}(t)$ be the position of its center of mass, and assume a diffusive dynamics for each component
\begin{eqnarray}
    \frac{d X(t)}{dt} = \sqrt{ 2 D_{T}}\  \xi_{T}(t)
    \label{eq_sub_T}
\end{eqnarray}
    with $\langle \xi_T(t) \rangle =0$, $\langle \xi_T(t) \xi_T(t') \rangle = \delta (t-t')$ and the translational diffusion coefficient $D_T$.

Assuming the substrate to be isotropic, its rotational motion is described by a single angular velocity vector ${\vec \Omega}(t)$.
Similarly to the translational motion, we assume a diffusive dynamics for each component
\begin{eqnarray}
\Omega (t) = \sqrt{2 D_{R}} \ \xi_{R}(t)
\label{eq_sub_R}
\end{eqnarray}
with  $\langle \xi_R(t) \rangle =0$, $\langle \xi_R(t) \xi_R(t') \rangle = \delta (t-t')$ and the rotational diffusion coefficient $D_R$.

To set a scale, we introduce a characteristic length $R_0$, which corresponds to the spatial size of the substrate. At the start of the simulation, we set the center of mass of the substrate to be the origin and place a pair of probes at random positions within a sphere with radius $c R_0$ centred at origin, where $c \simeq 1$ is a constant. At each time step, each probe initially performs its own fBm displacement, then is followed by the extrinsic displacements due to substrate translation and rotation. We employ Rodrigues' rotation formula to compute the rotational displacement from ${\vec \Omega}(t)$.

Our purpose in this paper is to see the regime, where the confinement effect is not relevant.
In Fig.~\ref{fig3}, we set $\alpha=0.5, K_{\alpha}=0.5, D_T = 0.1, D_R=10^{-3}$,  and $R_0 = 20$, $c=0.5$. Then, on the time scale shown in Fig.~\ref{fig3} (at least, up to $\tau = 10^2$), the typical displacement of each probe due to its intrinsic motion (calculated as the square root of MSD) is less than $R_0$, thus ensuring our aim.
Note that if the viscous resistance to the probe particle and that to the substrate were characterized by the same viscosity, the translational and rotational diffusion coefficients of the substrate are, respectively, $D_T = a/R_0$ and $D_R = (a/R_0)^3$, where Eqs~(\ref{eq_sub_T}) and~(\ref{eq_sub_R}) are made dimensionless with the unit length $a$ and time $\tau_0$. However, this is not likely the case, for instance, in the chromatin experiment, where the viscosity would be scale-dependent as is usually expected for systems with mesoscopic structures.

As shown in Fig.~\ref{fig3} in main text, the observed ${\rm MSD}^{o}(\tau)$ (and also ${\rm MSCV}^{(o)}(\tau)$ on rotating substrate) fails to capture the intrinsic probe mobioity. For sub-diffusional probe ($\alpha < 1$), the contribution of the substrate motion is negligible on short time scale, but it eventually dominates on longer time scale. The cross over time scale $\tau_c$ can be estimated by comparing the intrinsic probe displacement $\sim \tau^{\alpha}$ with the extrinsic substrate contribution $\sim D_T \tau$ or $\sim R_0^2 D_R \tau$ for translational or ratational contribution. One thus finds $\tau_c = \max \{\tau_{c,T}, \tau_{c,R}\}$ with $\tau_{c,T} \sim D_{T}^{1/(\alpha -1)}$, $\tau_{c,R} \sim (R_0^2 D_{R})^{1/(\alpha -1)}$. For probes with $\alpha=1$, such a crossover is not expected, where the motion of the substrate leads to a higher apparent diffusion coefficient.

\subsection{Model of polymer simulation}
\label{App2}
We employ a coarse-grained bead-spring model routinely adopted in literature, see Ref.~\cite{Michieletto2021} for details.
Briefly, we model each polymer chain as a linear sequence of $N$ beads of size $\sigma$. These beads are connected along the backbone by finite-extension nonlinear-elastic (FENE) bonds, and the excluded volume interactions between beads are imposed through the purely repulsive Lennard-Jones potential. We also introduce the bending potential $U_{bend}(\theta) = k_BT (l_p/a) ( 1 - \cos{\theta})$, where $\theta$ is the angle formed between consecutive bonds, and $l_p = 5 \ \sigma$ is the persistence length.
The polymer solution is composed of $M=109$ polymers with length $N=600$ in a cubic periodic box with side length $L=87 \ \sigma$ such that the monomer number concentration is $0.1\sigma^{-3}$. With this concentration, polymers with $N=600$ and $l_p=5 \  \sigma$ are well entangled. 

Molecular dynamics simulations are performed using the LAMMPS package~\cite{Plimpton1995}.
The position of each bead evolves according to the under-damped Langevin equation, where the bead mass $m$ and the friction coefficient $\gamma$ are set to satisfy $m/\gamma = \gamma \sigma^2/k_BT$.  The thermal noise satisfies the fluctuation dissipation theorem. To integrate the equations of motion, we employ a velocity-Verlet algorithm with time step $dt = 0.01 \gamma \sigma^2/k_BT$.
The length and time are measured in units of $\sigma$ and $\tau_{\sigma} \equiv \gamma \sigma^2/k_BT$, respectively.

Let $n \in [1,N]$ denote the bead label in each polymer. In the analysis of trajectory data, we adopt one monomer $n=100$ in each chain as tagged, and use trajectories of these tagged monomers to compute ${\rm MSD}$, ${\rm MSCD}$(inter) and ${\rm MSCD_c}$ in Fig.~\ref{fig6} and $\Delta^{NG}$, $\Delta^{NG}_d$(inter) and $\Delta^{NG}_{d,c}$ in Fig.~\ref{fig8}. For the intramolecular analysis, we select two monomers $(n_1, n_2) = (100, 500)$ in each chain as tagged to compute ${\rm MSCD}  {\rm (intra} \ m=400)$ in  Fig.~\ref{fig6} and $\Delta^{NG}_d {\rm (intra} \ m=400)$ in  Fig.~\ref{fig8}.  
 

\end{document}